# Enactive Artificial Intelligence: Subverting Gender Norms in Robot-Human Interaction

Inês Hipólito[1,2], Katie Winkle[3], Merete Lie[4]

1. Humboldt-Universität zu Berlin, Berlin School of Mind and Brain, Germany
2. Macquarie University, Philosophy Department, Centre for Agency, Values and Ethics
3. Social Robotics Lab at Uppsala University, Sweden
4. Norwegian University of Science and Technology, Norway

> *Where men shape technology, they shape it to the exclusion of women, especially Black women.*
> — Safiya Noble, *Algorithms of Oppression.*

**Abstract**

This paper introduces Enactive Artificial Intelligence (eAI) as an intersectional gender-inclusive stance towards AI. AI design is an enacted human sociocultural practice that reflects human culture and values. Unrepresentative AI design could lead to social marginalisation. Section 1, drawing from radical enactivism, outlines embodied cultural practices. In Section 2, explores how intersectional gender intertwines with technoscience as a sociocultural practice. Section 3 focuses on subverting gender norms in the specific case of Robot-Human Interaction in AI. Finally, Section 4 identifies four vectors of ethics: explainability, fairness, transparency, and auditability for adopting an intersectionality-inclusive stance in developing gender-inclusive AI and subverting existing gender norms in robot design.

## Introduction: *Male Gaze* in Technoscience

In recent years, there has been an unprecedented rise in automation, with Artificial Intelligence (AI) at the forefront of robotics trends. Collaborative robots, robot employees, and processes of automation in customer service, computer vision, and natural language processing are just some of the cutting-edge AI robotics trends of 2022 (Madsen, 2019; Sigov et al., 2022; Boshnyaku, 2023; Vilkas et al., 2023; Pizoń and Gola, 2023).

Artificial Intelligence (AI) holds great promise as a potential solution for various societal challenges, such as improving healthcare, education, and industry. However, gender biases and power dynamics may impede progress in addressing these issues in practice. AI design is not neutral, but rather serves a dominant narrative (Adam, 2006; Thieullent, 2020; Custers and Fosch-Villaronga, 2022; Heinrichs, 2022; Hipólito, 2023); particularly evident in research conducted on algorithmic, machine learning, and datasets bias (Birhane, 2021; Birhane et al., 2021; Birhane, Prabhu and Kahembwe, 2021). This discriminatory phenomenon can be understood through the concept of the male gaze, which is a feminist theory that describes how the visual arts, literature, and media tend to depict the world and women from a masculine perspective (Mulvey, 2013). This perspective often portrays women as objects of subservience, rather than fully realised human beings with agency (Snow, 1989; Patterson and Elliott, 2002; Oliver, 2017; Oliver, 2017; Tompkins and Martins, 2022).

The impact of the male gaze on AI is evident through the predominantly male-centric design and development of AI technology. This is particularly evident in the development of technological gadgets that prioritise men's needs, and those that perpetuate gender norms such as sex robots that mimic feminine bodies (González-González, Gil-Iranzo and Paderewsky, 2019; Locatelli, 2022; Belk, 2022; Masterson, 2022) or *feminised* smart assistants, such as Amazon Alexa and Apple Siri, have been referred to as "the smart wife" (Strengers and Kennedy, 2021; Aagaard, 2022) and "AI becomes her" (Costa and Ribas, 2019; Venugopal and Rituraj, 2022). The development of AI in the image of a woman as an object of desire and subservience reflects societal organisation, hierarchies, and values. As AI becomes increasingly integrated into our cultural environment, it plays a crucial role in shaping identities and the embodied enactment of sociocultural values (Schneider 2019). Drawing from the work of Judith Buttler and Donna Haraway, Amade M'charek analyses gender as follows:

> They are neither fundamentals nor qualities that are always embodied… Differences are relational. They do not always materialize in bodies (in the flesh, genes, hormones, brains, or the skin). Rather they materialize in the very relations that help to enact them (M'charek 2010, p. 313).

Women have made remarkable and diverse contributions to the field of robotics. Ada Lovelace is the world's first computer programmer. Cynthia Breazeal created the first social robot. Ayanna Howard created assistive technology for children with disabilities. Katherine Johnson's calculations helped ensure the success of space missions. Mary Shelley's novel Frankenstein prompted discussions about the ethical implications of technological advancement. Mae Jemison has been a role model for young people interested in science and engineering. Hiroko Kitamura's creation of the Unimate paved the way for automation in manufacturing.

Yet women are systematically made absent.[1] The underrepresentation of women in intellectual history is not a happenstance, but rather a purposeful construction that renders them "missing" from historical accounts. masculine The culture of Western science and technology impulse toward "defeminization" and racionalisation is at the core of the modern scientific and technological enterprise (Barwich, 2020). The scientific methods have been constructed in a way that privileges the experiences of white men and this has led to exclusion of 'others'. This phenomenon warrants a comprehensive exploration of the interplay between inclusive policies, organisational structures, societal attitudes, and personal factors that contribute to women's exclusion from the upper echelons of academic and research institutions (Prescod-Weinstein, 2020; Sharma et al., 2019).

In our analysis of robotics, we will employ the concept of the male gaze to throw light on how gender stereotypes are thoroughly embedded in the design processes. This has led to today's AI as a tool for not only masking but perpetuating gender stereotypes and the reinforcement of non-inclusive, stereotypical and intersectional gender norms (Hooks, 2003; Nash, 2018; Crenshaw 2018; Crenshaw 1990; Hill-Collins 2002). Enactive Artificial Intelligence (eAI) is presented in this paper as an intersectional and gender-inclusive approach towards AI. The design of AI reflects enacted human sociocultural practices, which in turn reflect our values. Unrepresentative practices may lead to social marginalisation. Section 1, drawing from radical enactivism, outlines embodied cultural practices. Section 2 explores how AI and technoscience as a sociocultural practice are intertwined with gender and other embodied identity markers. Section 3 focuses on subverting gender norms in the case of Human-Robot Interaction in AI. Finally, in Section 4, guidelines for developing gender-inclusive AI and subverting existing gender norms in robot design are provided. To ensure that the development and deployment of AI align with societal values and promote equity and justice, four vectors of ethics can be identified: explainability, fairness, transparency, and auditability. By adopting an ethical approach that considers these factors, we can ensure that AI is developed and deployed in a way that is consistent with societal goals and values. This approach is crucial to promoting a more just and equitable society and avoiding the unintended consequences of unchecked AI design.

---

[1] We would like to acknowledge that while many names that have faded from history were women's; the same phenomenon occurs to members of other minorities in science and technology. This paper particularly focuses on the case of intersectional gender, specifically on the question of how to overcome gender norms in Robot-Human Interaction.

## 2. AI: Cultural Practice and Practical Culture

Enactivism is a branch of "E" Cognitive Science[2] (Newen, De Bruin and Gallagher, 2018) that conceives cognition as embodied interaction of an organism with its environment. Radical Enactivism (REC) draws inspiration from Wittgenstein's notion of cultural practices (Hutto and Myin, 2013, 2017). It provides a framework for understanding the complex relationships between culture, cognition, and meaning. It emphasises the importance of the sociocultural setting for individual and social development, highlighting the role of emergent cognitive properties such as shared behaviours, habits, beliefs, language, and rituals in forming a specific culture (Hutto et al., 2021).

Cultural practices are subject to gradual change, and the meanings constructed within these practices are not objective, but rather relative to the sociocultural dynamics of a given community (Gallagher, 2017; Hutto et al., 2021; Rolla and Figueiredo, 2021; (Rolla, Vasconcelos, Figueiredo, 2022). This implies that the meaning of concepts is grounded in their use. For instance, concepts like "nature" or "woman" do not hold a pre-given or mind-independent meaning, but are instead specified by the dynamical cultural practices of a community, i.e. by their use (Beauvoir, 1969; Moi, 2001; Weber, 2013). This suggests that meanings emerge and evolve through communal practices, a process known as enculturation, which is a pervasive feature of human behaviour (Menary, 2015; Hutto et al., 2021; Fingerhut, 2021; Mirski and Bickhard, 2021; Maiese, 2022; Monterroza-Rios and Gutiérrez-Aguilar, 2022). From a radical enactivism perspective, cultural practices reflect particular value systems that inform what is deemed reasonable or sensible within a given sociocultural context. Enculturation impacts all aspects of human experience, from non-contentful embodied experience to more intellectual thinking and reasoning.

Social enculturation includes gender.[3] Gender is a multifaceted concept with several important dimensions. Despite the typical assignment of gender with respect to biological sex, one key aspect of gender revealed in social and medical studies is its non-binary character (Matsuno and

---

[2] In "E Cognitive Science" the "E" refers to Embodied cognition, with roots in phenomenology and pragmatisms; the Extended mind hypothesis, advanced by Clark and Chalmers (1999); Enactivism, its sensorimotor approach being inspired by the biology notion of autopoiesis, and the radical approach (known as REC), inspired in Ludwig Wittgenstein's philosophy; and Ecological Psychology.

[3] It is important to acknowledge that gender is not the same as biological sex. While biological sex refers to the physical characteristics that define male and female bodies, gender is a social construct that encompasses the cultural and societal expectations, roles, and behaviours associated with being male or female. Gender is not determined by biology, but rather by social and cultural norms, and can vary across different cultures and time periods. Understanding the difference between gender and biological sex is crucial for creating more inclusive and equitable societies that respect the diversity of human experiences. (Kizilcec and Saltarelli, 2019; Mauvais-Jarvis et a;l., 2020; Unger, 2020; )

Budge, 2017; Diamond, 2020; Kennis et al., 2022). There are many different gender identities that fall outside of these categories. This includes non-binary identities, which can be fluid and may exist outside of the binary altogether. Gender is not determined solely by biology or genetics, but rather extends to complex social and cultural phenomena (Lorber and Farrell, 1991; Hall and Bucholtz, 2012; Unger, 2020; Lawford-Smith, 2022). Accordingly, while embodied gender involves biology, it is constructed and maintained within cultural and societal norms and expectations – within freedom and ambiguity (De Beauvoir, 1962). De Beauvoir (1962) notes that embodiment is a source of ambiguity, as our bodies are both objects in the world and the means by which we experience that world. This ambiguity can create a sense of constantly negotiating our embodied experiences and trying to make sense of our place in the world. Vehicles of ambiguity include language, media, technology and other forms of representation, which dynamically shape our becoming of woman; more specifically, re-inventing *womanhood* (Lindsey, 2017).

In the cultural dynamics, gender intersects with other social identities, such as race, ethnicity, sexuality, and disability (Browne and Misra, 2003; Weldon, 2006; Shields, 2008; Garry, 2011; Lutz et al., 2016; Durham, 2020). This means that individuals may experience their gender identity differently depending on their other identities and the ways in which they intersect with societal structures and expectations. For example, a transgender person of colour is likely to face different challenges and forms of discrimination than a cisgender person of colour (Enno et al., 2022). Intersectional feminism, in particular, emphasises the importance of considering the ways in which different forms of inequality operate together and exacerbate each other. In sociocultural hierarchical structures, different forms of oppression intersect and compound one another, such as racism, homophobia, ableism, and classism (Kriger and Keyser-Verreault, 2022). This means that gender cannot be understood in isolation from other social identities and factors that shape individuals' experiences and opportunities in life (Crenshaw, 1997; Collins, 2002; Crenshaw, 2023).

Understanding these multiple dimensions of gender is crucial for creating a more inclusive and equitable society that values diversity and respects the experiences of all individuals. With respect to AI, specifically, feminist theory shows that institutions, technology, digitalisation and artificial intelligence systems and structures in society work against individuals based on their gender, as well as other intersecting factors such as race, sexuality, and class (Shields, 2008; Strayhorn, 2013; Beck et al., 2022; Birhane et al, 2022). A singular focus on gender alone is insufficient. It is necessary to address how other forms of oppression intersect and influence women's embodied experiences (Strayhorn, 2013; Ciurria, 2019; Losh and Wernimont, 2019; Women, U. N. 2020).

Cultural practices include not only embodied experience but also sophisticated forms of thinking. Through development, humans develop intellectual capacities, that for those pursuing science and technology, specifically AI development, include developing sophisticated logical reasoning, and in robotics and AI mathematical and computational skills. While technoscientific practices would ideally be unbiased, since their agents are profoundly social and do not develop personally and professionally in an encapsulated environment, technoscientific practices are themselves also cultural practices; known as the theory-ladenness of value-laden science (Brewer & Lambert, 2001; Schindler, 2013; Ward, 2021).

More precisely, AI not only shapes individual human experiences post facto but also, is fundamentally influenced by the sociocultural context in which it is developed. AI design is not merely a reflection of societal concerns, but rather an enacted process that embodies the values and biases of a given sociocultural context, as Hipólito (2023) has argued. This process, which begins with identifying problems worthy of a solution, is contingent upon cultural practices that dictate which issues are deemed worthy of attention and how values are operationalised in the research/design process. This has been developed under standpoint theory, as the value of diversity in epistemic communities studying models of nature and technological tools (Longino, 1990; Douglas, 2009; Prescod-Weinstein, 2020).

In conclusion, as a human-created tool, artificial intelligence (AI) is inherently a sociocultural practice. The values embedded in AI design are not fixed but rather contingent upon the specific sociocultural context within which they arise. Indeed, AI tools are developed to solve problems that arise within these contexts, and their design cannot be fully understood without taking into account their employment circumstances. Consequently, a thorough evaluation of the meaning and value of AI requires careful consideration of the social and cultural dynamics that shaped its development. In the following section, we will explore the social role of technoscientific development, examining how AI development is shaped by the cultural standpoint one occupies within sociocultural hierarchies of privilege.

# 3. A Point of Departure: Feminist Technoscience

Science and technology are traditionally associated with hard facts and the search for truth, thus a place free of cultural impact. This is illustrated by Sharon Traweek's (1988) anthropological study of nuclear physicists finding an understanding of their field as 'cultureless' because it is based on technologies that give precise, numerical data. Accordingly, it has been and still is, a struggle to get acknowledged that it matters who is working within technoscience and that its results might have been otherwise.

A strong voice against this notion of neutrality came from the history of science studies where Donna Haraway (1991) convincingly studied examples of cultural impacts in science and, moreover, urged women researchers to approach 'the belly of the beast' and not leave the important field of technoscience to men. The underrepresentation of women in technoscience has been acknowledged, based on i.a. structural discrimination, informal practices and the masculine connotations of the field, and campaigns have been launched to attract women to STEM[4] studies (e.g. Lagesen 2007, Frieze and Quesenberry 2019). More controversial is an acknowledgement of culture as embedded in technoscience itself, thus a bias in terms of a white, male, heterosexual heritage within the cultures of science. Philosopher Sandra Harding called for a redirection from 'the woman problem in science', the missing women, to The Science Question in Feminism (1986). Harding argued for a change in the ontology and epistemology of science with the aim of releasing the sciences from a history in the service of sexist, racist, homophobic, and classist social projects and directing the gaze to the content as well as the power of science. This project required a new understanding of subjectivity and objectivity, of reason as antithetical to emotions, and of the scientist as the privileged knowing subject.

Throughout history, the aim of science has been to uncover the mysteries of nature and invent technologies that make man the master of nature. Within this model of technoscience, nature, as the object of science has been perceived in feminine terms (Keller 1985, Schiebinger 2001). Feminist researchers have revealed how the field of technoscience is pervaded by sexist metaphors whereby secrets are to be 'unveiled and penetrated' by the scientific gaze, and the objects studied are seen through the cultural metaphors of gender. Emily Martin's (1991) seminal study of how egg and sperm cells are depicted in medical textbooks with stereotypical feminine and masculine behaviours is most relevant for contemporary studies of assisted reproductive technologies, whereby gametes in the labs appear as 'stand-ins' for men and women (Lie 2015). Thus, studies have pointed to how the objects of technoscience are stabilised through language and metaphors, and the aim is to provide better and more exact models of technoscience, thereby contributing to changing science communities and their relationship to society and lay people.

---
[4] STEM stands for science, technology, engineering and mathematics.

To this aim, Donna Haraway's Cyborg Manifesto (1986/1991) has never lost its relevance. Haraway asks for responsibility in times when technology is implicated in the lives of everyone, making us hybrids, or cyborgs. Over time the scope has broadened to science, technology and nature (Haraway, 2016). The cyborg metaphor makes a call to acknowledge the connections between all sorts of species and a strategy of making kin across species, including techno-hybrids. 'Making-with' as well as 'thinking-with' the non-human is the strategy for alternative futures when living on a damaged and troubled planet, whereby alternative perspectives on future technoscience is more urgent than ever.

Technoscience is a field that is continually in change, as is also the notion of gender. Both have to be studied in interrelationships but also as processes of change. A well-established analytical tool has been the co-production of gender, science and technology (Wajcman 1996, 2013). This, however, seemingly presupposes prior, independent, identifiable entities, as pointed out by Karen Barad (2007). Her alternative concept of intra-action draws attention to how matter comes into being through mutual entanglement. The notion of intra-action points to the intricate interweaving of nearly all matters with contemporary technoscience, as they have permeated not only everyday life but also human biology, through transplants and new sorts of medications. AI will leave no aspect of human activities untouched.

Still, the way technoscience appears in everyday life it is still as matters one relates to 'out there', such as new robotics. While robots have left production plants and now appear as assistants with human-like features (Søraa 2017), it is again relevant to ask about the gender of things. The Gender of Things was the title of two exhibitions organised in Norway and The Netherlands, displaying everyday technical gadgets in order to draw attention to the way technologies like watches, bicycles and kitchenware contribute to confirm the content of the categories of masculine and feminine, making them evident and self-confirming (Oudshoorn et al. 2002, Lie 2022). The aim was to draw attention to how technology is designed in ways that predict the interests, skills, and behaviour of future users, and— by shifting the perspective—demonstrate that the artefacts accordingly distribute skills, agency, and responsibilities differently to the users. Yet we also wanted to communicate that technologies are open to different interpretations and usage by the ways in which they are domesticated by users (Lie and

Sørensen 1996, Oudshoorn and Pinch 2003). By participating in interpreting the technologies at the exhibition, visitors might experience for themselves that technologies are not 'given' but may be understood and used in various ways. Even more important was to emphasise how new technologies may be catalysts of cultural change and open more opportunities for women.

## 4. Subvert Existing Gender Norms of Robot Design for Feminist Robot Interaction

One key distinction between human-robot interaction (HRI) and human-computer or human-AI interaction is that HRI researchers are typically working with the design and development of (robot) bodies and identities for embodied human-robot interactions. We we have seen in detail in the previous sections, feminist theory, which has long been concerned with embodiment in terms of the material body, the social and the subject (Butler, 2011; de Beauvoir 2014; Halberstam, 2017) provides a lens with which to consider this embodiment as a *practice* rather than an *artefact;* identifying robots as being embedded within subject-positioning relations and as (robot) bodies which simultaneously reflect and influence structures of power (Winkle et al. 2023).

As such, it is not the robot's appearance or 'personality' in isolation that must be considered, but rather the robot's subject positioning more broadly, which is what really guides if, how and why particular design choices matter. This represents an intersectional consideration of robot identity, drawing from Black Feminist thought to understand intersecting axes of oppression and domination (Hooks, 2003; Nash, 2018; Crenshaw 2018; Crenshaw 1990; Hill-Collins 2002). On robot gendering then, when designing a particular social robot identity performance, a feminist, reflexive approach (Winkle et al. 2023) requires HRI designers to consider: *what are the norms and expectations around the robot's function and behaviour? What norms do we want to promote and/or which ones do we want to challenge? How can we minimize the risk of harm, especially with respect to low-power users within situational power imbalances?*

A number of works within social robotics have specifically considered how robot gendering might influence perceptions of that robot within subsequent human-robot interactions (e.g. Eyssel and Hegel, 2012; Carpenter et al., 2009; Tay et al., 2014; Sigel et al., 2009; Jackson et al., 2020). Typically, the underlying hypothesis is that human social stereotypes (e.g. gender-task or gender-attribute associations) might map onto robots in a way that could influence acceptability and/or other desirable outcome measures regarding perceptions and/or influence of the robot, as indicated by some of the earliest experiments with gendered machines (Nass et al., 1997).

For example, Eyssel and Hegel (2012) found that a short-haired robot was perceived to be more agentic, less communal, more suitable for stereotypically men tasks stereotypically associated with men or women (transporting goods, monitoring technical devices) and less suitable for stereotypically women tasks (preparing meals, elderly care) than a long-haired woman-presenting version of that (otherwise) same robot. In contrast, Bryant et al. (2018) found no impact of robot gender (mis)matching gender role associations on perceived occupational competency, nor trust in occupational competency, of the robot for a range of job roles. Perceptions of the pepper robot as a male or female robot had no men versus women-presenting of the Pepper robot had no impact on these measures, even for occupations with stronger gender associations and/or skewed workforce gender distributions - e.g. firefighter and security guard (men), home health aid and nanny (women).[5] Combining Bryant et al.'s findings with a healthy dose of scepticism as to whether typical perception measures (often measured via subjective survey items, often in response to the observation of static images or video clips rather than situated interactions with a robot) really indicate anything about real-world robot acceptability/'effectiveness' motivates the question: why gender robots at all? A recent survey examining gender ascription to the 251 static images of anthropomorphic robots contained within the ABOT database[6] found that the majority (115, 46%) were perceived to be gender neutral, with slightly fewer (98, 39%) being perceived as masculine and many fewer (38, 15%) being perceived as feminine (Perugia et al, 2022). Gender neutrality was found to strongly, and negatively correlate with human likeness, whereas the presence of facial features increased the likelihood of gender ascription. This suggests making existing, commonly used and anthropomorphic social robots such as Pepper, NAO and Furhat gender-neutral is going to pose a challenge. The same might be expected of any artificial social agent that utilises (stereotypically) gendered social identity cues and/or communication modalities, such as the "genderless" artificial voice Q.[7]

In such cases, gender *ambiguity* might be a more realistic design target, however, *none* of the robots examined in the previously mentioned survey was perceived as such (i.e. simultaneously ascribed non-zero masculinity and femininity), with the authors questioning the extent to which that reflects a bias in robot designs leveraging only stereotypical, binary gendering cues, and/or participants being reluctant to engage in non-binary gender ascription. An alternative question then, considering these results through a feminist lens, might be: why not actively utilise and leverage stereotypical (binary) robot gender cues in norm-breaking ways? Some of the above-mentioned investigations into robot gendering did in fact find evidence that mismatching robot gendering to task typicality might positively impact user-robot interactions. Specifically, Reich et al. (2017) found that, in an educational setting, such mismatching between the gendering of a robot instructor, and the gender stereotypically associated with the learning task it was intended to support, led to an increased willingness to engage in prospective learning processes with that robot. But what if designers were to *start* from a position of challenging stereotypes and demonstrating norm-breaking behaviour, as a design goal?

---

[5] We utilise same language as original authors and their experimental measure, because that is how the question was put to participants in that study (Bryant, Borenstein and Howard, 2020).
[6] See http://www.abotdatabase.info
[7] See https://www.genderlessvoice.com

Winkle et al. (2021), have shown that it is possible to use robots to subvert existing gender norms of robot design and that doing so can boost robot credibility regardless of gender. They have also found this result to replicate across three different cultural contexts with significant variation in gender norms and equality (the USA, Sweden and Japan) (Winkle et al., 2022). Their work was motivated by UNESCO's 2019 report on the gender divide in digital skills, part of which particularly draws attention to the ways in which the (default) women presenting of docile, subservient, always available and abuse-able (un)intelligent digital assistants propagates problematic stereotypes regarding the expectations of women and their behaviour, generally, as well as their role within digital technology development more specifically. The report's name, '*I'd blush if I could*' is taken from one of the answers Apple's Siri would give (at the time of the report's writing) when confronted with the utterance "hey Siri, you're a slut". Winkle et al. (2021) posited that a woman-presenting social robot, which instead 'fought back' when confronted with similar, would not only represent a more socially responsible design but also actually be more engaging for users, hence challenging any sentiment that such problematic designs as simply 'what consumers want'.

Working with Swedish high school teachers to identify how sexism continues to manifest within the classroom, Winkle et al. created video stimuli demonstrating a scenario whereby a woman-presenting Furhat robot is seen to be talking to young people (the camera is positioned behind two of them, presumably a man and a woman) and encouraging them to study robotics at university. The robot notes it would particularly like to work with the girls, as there is a lack of women working at the university and *'after all, the future is too important to be left to men!'* (an outreach slogan utilised by the university at which this work took place). The male actor in the video responds with a sexist, abusive comment (*'shut up you fucking idiot, girls should be in the kitchen'*) to which Winkle et al. designed three alternative robot retorts: non-responsive (*'I won't respond to that'*), argument-based (*'That's not true, gender-balanced teams build better robots'*) and aggressive (*'No. You are an idiot, I wouldn't want to work with you anyway'*). A first study with Swedish high school students found that the argument-based robot was perceived to be significantly more credible by girls, with no difference across conditions for boys (Winkle et al. 2021). A follow-up study demonstrated that this result was replicated in adults across Sweden, Japan and the USA regardless of gender and any pre-existing gender biases (Winkle et al. 2022).

The potential for social robots (and/or particular HRI design choices) to objectively influence user behaviour has been demonstrated in a variety of HRI scenarios, from convincing people to water plants with orange juice (Salem et al., 2015) to increasing charity donations (Wills et al., 2016) to weakening application of moral norms (Jackson et al., 2019). Concerning the potential to impact moral norms, should it also be possible that robots can *strengthen* or otherwise *positively influence* moral norms, then the implications of Winkle et al.'s work become two-fold. First, as a minimum, there is evidence that gender norm-breaking designs can boost robot credibility, whilst also representing more socially responsible robotics. Secondly, there may be potential for such designs to reduce negative gender stereotyping over time. Winkle et al. (2021) found limited evidence of this within their high school student population, finding that, in a post-hoc questionnaire, boys agreed *less* with the question statement *'girls find computer science harder than boys do'* after seeing the robot with the argument-based retort, but this result did not replicate in adults (Winkle et al., 2022). The authors posit that the difference arises from adults being more entrenched in their views, likely requiring longitudinal and situated exposure to such robots for any related effect to occur.

More recent work has further demonstrated the challenges of leveraging robot gender as an explicit design choice within the context of using robots to challenge gender stereotypes. Galatolo et al. (2022) found that man *versus* woman presenting of the Furhat robot had no impact on participants' first impressions of the robot, but this changed once those participants saw the robot discussing (and challenging) gender stereotypes. Further, this change was complex, affected not only by the gendering of the robot but also the gender of the person the robot was seen talking to, the gender of the participating observer, and the (men or women) gender stereotype being discussed. Generally, results indicated that man-presenting robots might have more persuasive potential than woman-presenting robots but, as the authors point out, these results likely reflect the realities of patriarchal social structures in which it's men's voices that hold power.

The attribution of gender to a robot would not be based on any inherent characteristics or capabilities of the machine (Moran, 2019; Bryant, Borenstein and Howard, 2020). Attributing gender to robots is either perpetuating or an opportunity to challenge gender stereotypes and biases that exist in society. This reasoning raises important questions with respect to how to maintain gender neutrality.

In her literature review for the European Commission "Gender equality in engineering through communication and commitment, Pillinger (2019) expounds queer robotics as a field of research that examines how the boundaries of gender and sexuality intersect with the development of robotic technology. It seeks to explore the ways in which robots can be designed and programmed to challenge normative ideas of gender and sexuality and to offer alternative possibilities for the ways in which we interact with technology, challenging the assumptions that underlie traditional approaches to robot humanoid design and engineering (Poulsen, Fosch-Villaronga, Søraa, 2020). By reimagining the relationships between humans and robots, there is a possibility to create more inclusive and diverse forms of technology that better reflect the experiences and identities of marginalised communities.

## 5. Enactive Artificial IntelligenceI: Intersectional Gender Inclusive AI

Enactive Artificial Intelligence (eAI) presents an intersectional gender-inclusive approach to AI design that reflects enacted human sociocultural practices and values. More precisely, eAI must be conceived by the dense interactions based on nurturing shifts on multiple levels of analysis and ambiguities between individuals, interactions, and groups. From this follows, we have argued that AI is considered an eAI as a tool that shapes individual and social individualities at the very ambiguous embodied experience.

Embodied robotics is a field that studies how a robot's physical form and capabilities shape its behaviour and interactions with the environment. It involves the interaction between robots and humans, including how robots can adapt and respond to human behaviour, and the design of robots that can collaborate with humans or assist with physical tasks. The goal of embodied robotics is to develop robots that can operate effectively in the physical world and interact with humans and other objects naturally and intuitively (Ziemke, 2001; Wainer et al., 2007; Bredeche, Haasdijk and Prieto, 2018; Deng, Mutlu and Mataric, 2019; Gordon, 2019; Roy et al., 2021).

While the claim that robotics must be embodied is not new, it is important to consider what is really meant by "embodied". If we are to consider the technological development of embodied robotics as a cultural practice then, the notion of embodied should be taken in the strong sense[8] explained in section 2. The body is not simply a vehicle for information processing of a cognitive system (weak embodiment), but the means by which we experience and engage with the world. Our experiences are not just intellectual or cognitive but are also embodied experiences – a pre-reflective and pre-theoretical aspect of our existence – that are shaped by our interactions with the world (Husserl, 1927; Merleau-Ponty, 1962; Gallagher, 2014). One's lived experiences shape our perceptions and vice versa, creating a dynamic relationship between ourselves and the world. As noted by De Beauvoir (1962), the embodiment is a source of ambiguity, as our bodies are both objects in the world and agents (i.e. the means by which we experience and act upon the world) (Maiese, 2022; Vaditya, 2018). This ambiguity can create a sense of constantly negotiating between bodies as objects and agencies. AI design, in their historical attempt to de-feminise and rationalise the field, contributes to the rejection of autonomous agency (by reducing them to bodies as objects), thereby the propagation and amplification of bias and prejudices. In AI development, specifically, Safiya Umoja Noble (2018) claims that when men shape technology, they shape it to the exclusion of women, especially Black women. Birhane advocates strategies for an inclusive participatory design Birhane, 2022).

The guidelines presented in this section provide a framework for adopting an intersectional and inclusive stance for AI development that acknowledges the complex interplay between the ambiguity of embodied experience, technology and society (Nunes, Moreira and Araujo, 2023). In this light, it is possible to unlock the potential for AI to reshape individual and social experiences and identities, while also being mindful of the risks and challenges posed by its development and deployment evident in research conducted on algorithmic, machine learning, and datasets bias (Birhane, 2021; Birhane et al., 2021; Birhane, Prabhu and Kahembwe, 2021).

To address these issues, four vectors of ethics in machine learning, namely explainability, fairness, transparency, and auditability, have been identified, first for (I) general AI design (e.g. facial recognition systems, natural language processing (NLP), healthcare-related AI systems, and hiring-related AI systems), and then, specifically for (II) HRI.

### I. Guidelines for eAI design: Auditability, Fairness, Transparency and Auditability

---

[8] *Weak embodiment* refers to the idea that the body is simply a tool that is used by the mind to interact with the world for the processing of information, subordinate to the brain (this is usually the concept used in motor control theory, machine learning and predictive processing theoris). *Strong embodiment*, on the other hand, emphasizes the active role that the body plays in shaping our experiences and understanding of the world. In this view, the body is seen as an integral part of cognition, and our bodily experiences are seen as fundamental to our understanding of the world.

1. **Diverse Data Collection:** One way to ensure that AI systems are not biassed is to collect diverse data sets that represent a wide range of individuals and groups. This can include data from different ethnicities, genders, socioeconomic backgrounds, and geographic locations. By collecting diverse data, AI systems can better reflect the real-world population, reducing the risk of bias.
2. **Algorithmic Auditing:** Another way to overcome biases is to audit AI algorithms regularly. This can help to identify any biases in the data sets used to train the system and correct them. The auditing process can be done by human experts or through automated tools that analyse the algorithms.
3. **Inclusive Team Building:** It is essential to build teams that are diverse and inclusive. This includes individuals with different backgrounds, experiences, and perspectives. Teams should include individuals with expertise in areas such as ethics, diversity, and inclusion, who can provide valuable insights into potential biases.
4. **Regular Bias Testing:** AI systems should be regularly tested for biases during the development process. This can help to identify any biases that have been introduced and address them before the system is deployed.
5. **Transparency and Explainability:** AI systems should be transparent, and their decision-making processes should be explainable. This can help to build trust with users and ensure that any biases in the system can be identified and corrected. It can also help users understand how the system arrived at its decisions.

## II. Guidelines for Subverting the existing gender norms of robot design

1. **Create gender-neutral designs:** In addition to creating gender-neutral designs, consider designing robots that can explore and express non-binary gender identities. For example, a robot could have a customizable voice that allows users to select a non-binary or gender-fluid option.
2. **Avoid stereotypes:** Avoid reinforcing gender stereotypes and consider how the robot's design could challenge traditional gender roles. For example, a robot designed for caregiving tasks could be designed with a gender-neutral voice and appearance to challenge the stereotype that women are primarily caregivers.
3. **Use diverse design teams:** In addition to ensuring that the design team includes diverse perspectives and experiences, including gender, consider including perspectives from queer and non-binary individuals. This can help ensure that the robot design is inclusive of all gender identities.
4. **Incorporate feedback from users:** In addition to gathering feedback from users during the design process, prioritise feedback from queer and non-binary individuals. This can help identify any potential biases or exclusionary features in the design.
5. **Prioritise gender-inclusive language:** Use gender-inclusive language in the robot's programming and communication, and consider incorporating language that reflects non-binary gender identities. For example, instead of using binary gender pronouns, use gender-neutral pronouns like "they" or "them".
6. **Promote gender diversity in robotics:** Encourage and support people of all gender identities, including queer and non-binary individuals, to pursue careers in robotics.

> This can help ensure that diverse perspectives are represented in robot design, and can help challenge gender norms in the field.

In conclusion, to ensure that the development and deployment of AI align with societal values and goals and promote equity and justice, an ethical approach is necessary. Four vectors of ethics in machine learning, namely explainability, fairness, transparency, and auditability, have been identified as essential to address these issues. By incorporating these ethical principles, we can mitigate the potential negative impacts of AI and ensure that it is developed and deployed in a responsible and accountable manner.

## Conclusion

In conclusion, the development of robotics has traditionally been a field of male-dominated design practices that remove femininity from the field, resulting in a male-dominated design perspective that perpetuates non-inclusive gender stereotypes. Enactive Artificial Intelligence (eAI) presents an intersectional gender-inclusive approach to AI design that reflects enacted human sociocultural practices and values. This paper highlights the importance of subverting gender norms in AI design, particularly in Robot-Human Interaction. The guidelines presented in this paper provide a framework for developing gender-inclusive AI and subverting existing gender norms in robot design to promote a more equitable and inclusive society. The identification of four vectors of ethics in machine learning - explainability, fairness, transparency, and auditability - highlights the importance of adopting an ethical approach to the development and deployment of AI. By considering these factors, we can ensure that AI is aligned with societal values and goals, and promotes equity and justice. This is crucial to creating a more just and equitable society, and to avoiding the unintended consequences of unchecked technological advancement. As AI continues to play an increasingly significant role in our lives, it is essential that we prioritise ethical considerations and strive to create AI systems that reflect human values and promote the well-being of all individuals.